\pdfoutput=1

\documentclass{article}

\usepackage{PRIMEarxiv}

\usepackage[utf8]{inputenc} % allow utf-8 input
\usepackage[T1]{fontenc}    % use 8-bit T1 fonts
\usepackage{hyperref}       % hyperlinks
\usepackage{url}            % simple URL typesetting
\usepackage{booktabs}       % professional-quality tables
\usepackage{amsfonts}       % blackboard math symbols
\usepackage{nicefrac}       % compact symbols for 1/2, etc.
\usepackage{microtype}      % microtypography
\usepackage{lipsum}
\usepackage{fancyhdr}       % header
\usepackage{graphicx}       % graphics
\usepackage{multirow}
\usepackage{array}
\usepackage{algorithm}
\usepackage{algorithmicx}
\usepackage{algpseudocode}
\usepackage{amsmath}
\usepackage{amssymb}
\usepackage{mathtools}
\usepackage{enumerate}

\graphicspath{{/}}     % organize your images and other figures under media/ folder

\title{An Interpretable Generalization Mechanism for Accurately Detecting Anomaly and Identifying Networking Intrusion Techniques
}

\author{
  Hao-Ting Pai \\
  Bachelor Program of Big Data Applications in Business \\
  National Pingtung University \\
  \texttt{haotingpai@gmail.com} \\
   \And
  Yu-Hsuan Kang \\
  International Graduate School of Artificial Intelligence \\
  National Yunlin University of Science and Technology \\
  \texttt{M11063002@yuntech.edu.tw} \\
  \AND
  Wen-Cheng Chung \\
  Bachelor Program of Artificial Intelligence \\
  National Yunlin University of Science and Technology \\
  \texttt{sukoyachung@gmail.com} \\
}

\begin{document}
\maketitle

\begin{abstract}
The increasing complexity of modern network environments presents formidable challenges to Intrusion Detection Systems (IDS) in effectively mitigating cyber-attacks. Recent advancements in IDS research, integrating Explainable AI (XAI) methodologies, have led to notable improvements in system performance via precise feature selection. However, a thorough understanding of cyber-attacks requires inherently explainable decision-making processes within IDS. In this paper, we present the Interpretable Generalization Mechanism (IG), poised to revolutionize IDS capabilities. IG discerns coherent patterns, making it interpretable in distinguishing between normal and anomalous network traffic. Further, the synthesis of coherent patterns sheds light on intricate intrusion pathways, providing essential insights for cybersecurity forensics. By experiments with real-world datasets NSL-KDD, UNSW-NB15, and UKM-IDS20, IG is accurate even at a low ratio of training-to-test. With 10\%-to-90\%, IG achieves Precision (PRE)=0.93, Recall (REC)=0.94, and Area Under Curve (AUC)=0.94 in NSL-KDD; PRE=0.98, REC=0.99, and AUC=0.99 in UNSW-NB15; and PRE=0.98, REC=0.98, and AUC=0.99 in UKM-IDS20. Notably, in UNSW-NB15, IG achieves REC=1.0 and at least PRE=0.98 since 40\%-to-60\%; in UKM-IDS20, IG achieves REC=1.0 and at least PRE=0.88 since 20\%-to-80\%. Importantly, in UKM-IDS20, IG successfully identifies all three anomalous instances without prior exposure, demonstrating its generalization capabilities. These results and inferences are reproducible. In sum, IG showcases superior generalization by consistently performing well across diverse datasets and training-to-test ratios (from 10\%-to-90\% to 90\%-to-10\%), and excels in identifying novel anomalies without prior exposure. Its interpretability is enhanced by coherent evidence that accurately distinguishes both normal and anomalous activities, significantly improving detection accuracy and reducing false alarms, thereby strengthening IDS reliability and trustworthiness.
\end{abstract}

% keywords can be removed

\section{Introduction}
The Federal Bureau of Investigation (FBI) reports a 7\% increase in cybercrime complaints in the U.S., with expected losses surpassing \$6.9 billion \cite{FBI2021}. Globally, Barron’s estimates cybercrime costs to reach an astounding \$6 trillion \cite{Presse2022}. To combat these rising cyber threats, Intrusion Detection Systems (IDS) have been developed \cite{gu2021effective}, \cite{moustafa2019outlier}, \cite{odiathevar2021online}. However, traditional IDS often lack transparency, hindering administrators and security experts from fully understanding the automated decision-making processes.

Explainable AI (XAI) techniques \cite{dwivedi2023explainable}, offering explanations for predictive results, enhance trust and confidence in AI systems. Applying XAI to IDS presents challenges, requiring further research for the effective identification of attack surfaces and the interpretable justification of model outputs \cite{moustafa2023explainable}. For a comprehensive understanding of the decision-making process, it is imperative to link interpretations of the model’s predictions with relevant data sources, identified attack surfaces, and specific events. While interpretable methods like linear models, decision trees, and generalized additive models offer insights into predictions through linear combinations of feature values, they often fall short in accuracy compared to complex models like neural networks.

This paper elaborates on the Interpretable Generalization Mechanism (IG), an innovative approach primed to revolutionize IDS capabilities, thereby expanding the frontiers of cyber security. Addressing the rising complexity and sophistication of cyber threats, IG adopts an advanced yet comprehensible approach to attack recognition. IG leverages coherent pattern recognition to effectively differentiate between normal and anomalous network traffic, a crucial aspect for the early detection and mitigation of cyber threats. 

The generalization capability of IG, evidenced through extensive testing on datasets like NSL-KDD, UNSW-NB15, and UKM-IDS20, is notable, particularly with low training-to-test ratios. For example, in NSL-KDD, IG achieved an AUC of 0.94 with a 10\%-to-90\% ratio, showing similar high performance in other datasets. This robustness across varied datasets and scenarios underscores IG's adaptability and reliability, especially in identifying anomalous instances without prior training exposure, which is crucial for addressing dynamic cyber threats. 

Based on coherent patterns, IG facilitates transparent, interpretable decision-making, essential in forensics \cite{kuppa2021adversarial}, thereby fostering understanding and trust among security experts and administrators. Furthermore, the reproducibility of IG in diverse settings not only reinforces its reliability but also bolsters confidence in its applicability to real-world scenarios. By circumventing the need for extensive fine-tuning and reducing randomness, IG ensures a decision-making process that is both transparent and interpretable. This capability, vital in forensic analysis, enables security experts and system administrators to grasp the underlying rationale behind each decision rendered by the IDS. IG stands out as an invaluable asset in combating cyber threats due to its accuracy, interpretability, and transparency.

\section{Related Work}

\subsection{A Machine-learning Based IDS}
Machine learning has been effectively utilized in IDS. Notable examples include the integration of naive Bayes feature embedding with Support Vector Machine (SVM) \cite{gu2021effective}, the development of the Outlier Dirichlet Mixture-based Anomaly Detection System (ODM-ADS) \cite{moustafa2019outlier}, and the application of Deep Neural Networks (DNN) informed by Simulated Annealing Algorithms (SAA) and Improved Genetic Algorithms (IGA)\cite{chiba2019intelligent}. Moreover, the exploration of Wrapper Based Feature Extraction Units (WFEU) in conjunction with Feed-Forward Deep Neural Networks (FFDNN) has been reported \cite{kasongo2020deep}. To assess their efficacy, these studies have employed datasets such as NSL-KDD, UNSW-NB15, and UKM-IDS20. A summary of the data processing techniques, feature sets, and train-test ratios used in these studies is provided in Table \ref{tab:table1}. The results demonstrate high accuracy across these methods, with each employing a distinct approach to feature engineering during data preprocessing. Indeed, feature engineering is pivotal in the context of Explainable AI (XAI) as it substantially influences the interpretability and transparency of machine learning models. In the development of IG, our approach emphasizes preserving the original data structure, deliberately avoiding techniques such as the removal of irrelevant features, feature selection, and resampling for balance.

\subsection{Why Should We Need an Interpretable-generalization IDS?}
Recent advances in machine learning-based IDS have emphasized enhancing explainability. Techniques such as Neural Attention Models \cite{andresini2022roulette}, Shapley Additive exPlanations (SHAP) \cite{oseni2022explainable}, and Local Interpretable Model-agnostic Explanations (LIME) \cite{tcydenova2021detection} have been increasingly applied. A significant study \cite{keshk2023explainable} utilized XAI methods to identify critical features, thus reducing runtime and enhancing accuracy while providing clearer insights into the correlation between features and attacks. Nonetheless, to fully comprehend intrusion techniques, the core of an IDS must inherently be interpretable. 

Interpretable models provide essential insights into AI decision-making processes \cite{linkov2020cybertrust}. While models like linear models, decision trees, and generalized additive models offer enhanced interpretability, they often fall short in predictive accuracy when compared to more complex, opaque 'black box' models. To address this, an innovative approach, Transparent Classification (TC) \cite{pai2022explainable}, has been introduced, aiming to strike a balance between interpretability and accuracy. Experiments with two datasets “Contraceptive Method Choice” and “Breast Cancer Wisconsin” demonstrated that TC obtains perfect accuracy of recall, but lower accuracy of precision. Notably, TC faces a high false alarm rate. According to the third rule, TC optimizes the model by classifying an instance as positive only when it has a zero score for both positive and negative classifications. Under these conditions, the model fails to provide actionable insights for cybersecurity forensics, even though it successfully performs classification. Therefore, achieving a balance between interpretability and low rates of false positives and negatives in IDS remains a formidable challenge.

\setlength{\extrarowheight}{2pt}
\begin{table}
 \caption{Overview of Data Processing Techniques, Feature Sets, and Train-Test Ratios in the Literature}
  \centering
  \begin{tabular}{p{0.5cm}p{2.5cm}p{4.5cm}p{5.7cm}}
    \toprule
    Ref.     & Datasets     & Features     & Instances (Normal, Attack)\\
    \midrule
    \multirow{2}{*}{\cite{gu2021effective}} & NSL-KDD  & 41 (All) & 125973 (53\%, 47\%) \\
    \cline{2-4} % 仅在2到4列之间画线
    & UNSW-NB15  & 47 (All) & 118000 (47\%, 53\%) \\
    \hline
    \cite{chiba2019intelligent} & NSL-KDD  & 12 from 41 & 148517 (52\%, 48\%) \\
    \hline
    \cite{gottwalt2019corrcorr} & UNSW-NB15  & 15 (5-19) from 41 & First 200,000 (unknown) \\
    \hline
    \cite{patil2019designing} & UNSW-NB15  & 3 (18, 23, 26) from 47 & 2540044 (unknown) \\
    \hline
    \cite{zhang2020model} & UNSW-NB15  & 20 from 47 & 5576 (45\%, 55\%) \\
    \hline
    \cite{mohammad2020improved} & UNSW-NB15  & 8 from 47 & 32755 (50\%, 50\%) \\
    \hline
    \cite{aksu2022mga} & UNSW-NB15  & 7 from 47 & 20000 (50\%, 50\%) \\
    \hline
    \multirow{2}{*}{\cite{wang2021intrusion}} & NSL-KDD  & 41 (All) & 47600 (48\%, 52\%) \\
    \cline{2-4} % 仅在2到4列之间画线
    & UNSW-NB15  & 47 (All) & 257632 (36\%, 64\%)  \\
    \hline
    \cite{al2021adaptive} & UKM-IDS20  & ARP poisoning: 2 from 48, DoS: 42 from 48, Scans: 40 from 48, Exploits: 40 from 48 & Total: 12887 (69\%, 31\%); ARP poisoning: 8909, 592; DoS: 8909, 1742; Scans: 8909, 597; Exploits: 8909, 1047 \\
    \hline
    \multicolumn{4}{p{14.8cm}}{\textbf{Data Preprocessing:} Naïve Bayes feature embedding \cite{gu2021effective}, A modified Kolmogorov-Smirnov Correlation Based Filter Algorithm \cite{chiba2019intelligent}, PCA and Pearson class label correlation \cite{gottwalt2019corrcorr}, Binary Bat Algorithm with Feature Similarity-based Fitness Function and Classifier Accuracy based Fitness Function (BBA+FSFF+CAFF) \cite{patil2019designing}, Genetic Algorithms (GA) \cite{zhang2020model}, The correlation-based Feature Selection technique \cite{mohammad2020improved}, Modified Genetic Algorithm (MGA) m-feature selection \cite{aksu2022mga}, Denoising AutoEncoder \cite{wang2021intrusion}.} \\
    \bottomrule
  \end{tabular}
  \label{tab:table1}
\end{table}

\begin{figure}
  \centering
  \includegraphics[width=16.5cm]{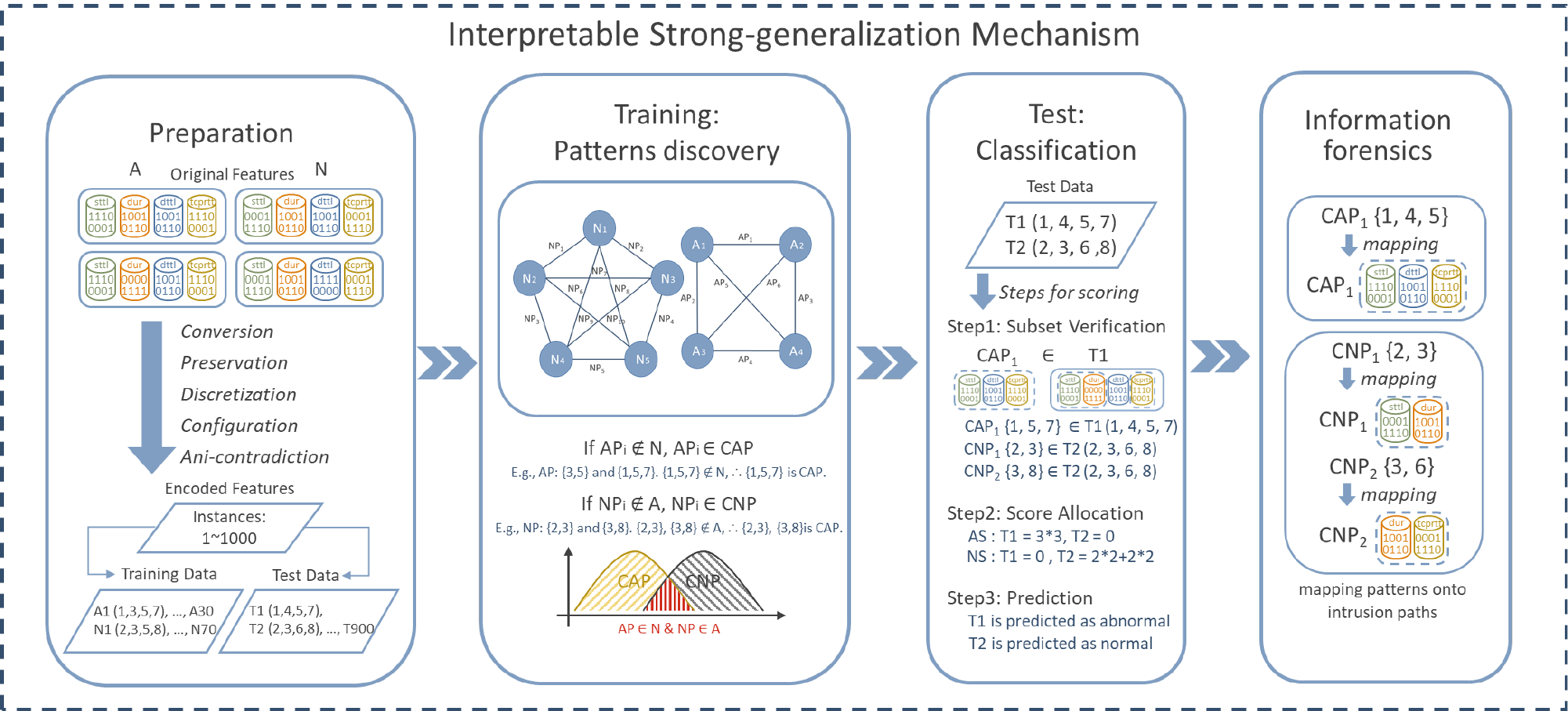}
  \caption{The Framework of Interpretable Generalization Mechanism.}
  \label{fig:fig1}
\end{figure}

\subsection{IDS with Forensics}
An advanced Intrusion Detection System (IDS) equipped with forensic capabilities marks a significant advancement in cybersecurity defense \cite{zipperle2022provenance}. Such systems not only monitor network activities in real-time but also capture comprehensive data regarding detected incidents. This facilitates in-depth post-incident analysis, assisting in the reconstruction of events leading up to an intrusion. A forensics-enabled IDS is crucial in identifying attack vectors and understanding the actions of intruders, shedding light on the extent of the breach and the specific actions of the intruders. Recently, XAI has been employed to investigate the characteristics of attacks. The study \cite{keshk2023explainable} revealed that activities with two features, proto = 0.5 (TCP) and service = 0 (none), are highly indicative of DDoS attacks. Through analysis, we found 141,523 instances with a proto value of 0.5 (TCP), of which 9,991 were DDoS attacks and 131,532 were normal. Similarly, of the instances with a service value of 0 (none), 13,489 experienced DDoS attacks, while 167,977 were normal. Depending solely on a single feature to classify an instance as normal or an attack can lead to a high false alarm rate, indicating that these features alone are insufficient for accurate DDoS attack identification. Therefore, when developing new cybersecurity defense methods, it is imperative to consider comprehensive attack paths, not only for identifying attacks but also for minimizing false alarms.

\section{Methods}
Fig. \ref{fig:fig1} presents the process of IG. In data preprocessing, we propose a reproducible manner to handle data issues such as varying types, identical instances with different labels, and missing values. In training, we establish coherent patterns that are occurred in only one class label, which are used for distinguishing normal and anomalous instances in the test. In test, for an instance, we calculate its normal and anomalous scores by using coherent patterns. Then, three regulations are utilized to determine whether an instance is normal or anomalous. To ensure comprehensiveness, the data are divided using a range of ratios, from low training-to-test ratios to high ones. In information forensics, we provide highlights of frequent intrusion techniques in different ratios of training-to-test.

\subsection{Data Preprocessing}
As Algorithm 1, five steps are involved, i.e., conversion, preservation, discretization, configuration, and anti-contradiction. In \textit{conversion}, we aim at identifying normal and anomalous instances instead of predicting its type of attack. Thus, various attacks are labeled as anomaly. In \textit{preservation}, we replace missing values with categorical codes rather than filling them with average values or other substitutes.  In \textit{discretization}, we use the z-score method to transform numerical values into categorical ones. In \textit{configuration}, we encode the columns to avoid errors of different columns but identical values, e.g., 1,1,1 becomes (0|1, 1|1, 2|1). In \textit{anti-contradiction}, we exclude instances that possess identical values yet are assigned different class labels, as outlined in Equation (1).
\begin{equation}
I_{\text{filtered}} = I - \{ x \in I \,|\, \exists y \in I \,:\, (x = y) \land (ans(x) \neq ans(y)) \}
\end{equation}
\begin{algorithm}
    \caption{Dataset preprocessing} 
    \begin{algorithmic}[1]
        \Statex \textbf{Input:} Cybersecurity dataset
        \Statex \textbf{Output:} Preprocessing cybersecurity dataset
        \State $I \leftarrow$ Cybersecurity dataset
        \State Replace missing values in $I$ with 'NotNumber'
        \State $ans \leftarrow$ Class labels from $I$
		\For {$i \coloneqq 0$ \textbf{to} $I.columns$ }
			\If {$I_{i}  = numerical $}
				\State $I_{i} \leftarrow \textsc{Zscore} (I_{i})$
			\EndIf
        \EndFor
        \State $I_{\text{filtered}} = I - \{ x \in I \,|\, \exists y \in I \,:\, (x = y) \land (ans(x) \neq ans(y)) \}$
        \State $M = \{ x \in I_{\text{filtered}} \,|\, ans(x) = \text{"Anomalous"} \}$
        \State $N = \{ x \in I_{\text{filtered}} \,|\, ans(x) = \text{"Normal"} \}$
        	\end{algorithmic} 
\end{algorithm}

\subsection{Training}
We split data into training and test sets. Let the total number of instances in the dataset be \textit{K}. Out of these, \textit{N} instances are normal, and \textit{M} instances are anomalous. Then \textit{K} is split into training and test sets based on a percentage ratio. Let $TR_{ratio}$ represents the training ratio, which varies from 10\% to 90\% in increments of 10\%. Consequently, the test ratio  $TE_{ratio} = 100\% - TR_{ratio}$. For the training set at any given $TR_{ratio}$, the number of normal instances $N_{train}$ and anomalous instances $M_{train}$ are calculated as follows: $N_{train} = N \times TR_{ratio}$ and $M_{train} = M \times TR_{ratio}$. Similarly, for the test set: $N_{test} = N \times TE_{ratio}$ and $M_{test} = M \times TE_{ratio}$.

\begin{algorithm}
    \caption{Identify distinguishable patterns and score} 
    \begin{algorithmic}[1]
        \Statex \textbf{Input:} Set of anomalous training instances $M_{train}$, set of normal training instances $N_{train}$
        \Statex \textbf{Output:} Distinguishable patterns and score
        \State {\bf Initialization:} $\text{CNP} \leftarrow \varnothing,\; \text{CAP} \leftarrow \varnothing,\; $
		\For {$\text{each pair of instances} \;i, j \in N_{train}$ }
            \State $\text{NP} \leftarrow N_{i} \cap N_{j}$
			\If {$\text{NP} \notin M_{train} $}
				\State $\text{Add} \; \text{NP} \; \text{to} \; \text{CNP}$
			\EndIf
        \EndFor
		\For {$\text{each pair of instances} \;x, y \in M_{train}$ }
            \State $\text{AP} \leftarrow M_{x} \cap M_{y}$
			\If {$\text{AP} \notin N_{train} $}
				\State $\text{Add} \; \text{AP} \; \text{to} \; \text{CAP}$
			\EndIf
        \EndFor
	\end{algorithmic} 
\end{algorithm}
As Algorithm 2, we discover coherent patterns by the following steps.
\begin{enumerate}[Step 1:]
\item Intersection of normal instances to generate normal patterns (NP).\\
We perform pairwise intersections to generate normal patterns. The intersection operation $N_{i} \cap N_{j}$ (where $i \neq j$) produces a set of common elements between any two normal instances, which is represented as Equation (2). Given $N_1 = (a, b, d, e)$, $N_2 = (a, b, e, f)$, $N_3 = (a, b, d, f)$, and $N_4 = (a, b, d, e)$, the intersection $N_{1} \cap N_{2}$ yields the normal pattern $N_{1-2} = (a, b, e)$. Moreover, $N_{train}$ themselves are normal patterns. Subsequently, we quantify the frequency of each normal pattern, e.g., $(a, b, e)\_2$, $(a, b, d)\_2$, $(a, b, f)\_1$, $(a, b, d, e)\_2$, $(a, b, e, f)\_1$, and $(a, b, d, f)\_1$.
\begin{equation}
 N_{i} \cap N_{j} = N_{i-j} = \{x | x \in N_{i} \  \text{and} \  x \in N_{j} \}
\end{equation}
\item Intersection of anomalous instances to generate anomalous patterns (AP).\\
We do the same as step 1. Given  $M_1 = (a, b, c, d)$, $M_2 = (a, b, d, g)$, and $M_3 = (b, c, d, g)$, we count the occurrence as $(a, b, d)\_1$, $(b, c, d)\_1$, $(b, d, g)\_1$, $(a, b, c, d)\_1$, $(a, b, d, g)\_1$, and $(b, c, d, g)\_1$.
\item Derivation of coherent normal patterns (CNP)\\
In Equation (3), we define CNP as those normal patterns that are not exactly replicated in the anomalous instances. In this case, we obtain three CNP: $(a, b, e)_2$, $(a, b, d, e)_1$, and $(a, b, f)_1$. The NP $(a, b, d)_2$ is excluded because it is a subset of an instance in $M_{train}$.
\begin{equation}
 \text{CNP} = \{N_{i-j} \mid N_{i-j} \notin \{M_{train}\}\}
\end{equation}
\item Derivation of coherent anomaly patterns (CAP)\\
In Equation (4), CAP are defined as those anomalous patterns that do not have an identical counterpart in the normal instances. This case has five CAP: $(b, c, d)\_1$, $(b, d, g)\_1$, $(a, b, c, d)\_1$, $(a, b, d, g)\_1$, and $(b, c, d, g)\_1$.
\begin{equation}
 \text{CAP} = \{M_{i-j} \mid M_{i-j} \notin \{N_{train}\}\}
\end{equation}
\end{enumerate}

\subsection{Test}
We calculate whether a CNP or CAP is a subset of a test instance, which then informs the scoring and classification process. As Equation (5), A pattern P (either CNP or CAP) is considered a subset of an instance $T$ if every element of $P$ is contained in $T$.
\begin{equation}
 P \subseteq T \Leftrightarrow \forall \kappa \in P, \kappa \in T
\end{equation}

\begin{enumerate}[Step 1:]
\item Normal score calculation. \\
For each test instance $T_{i}$, we calculate the normal score based on the occurrence and length of any CNP that is a subset of $T_{i}$. The normal-score $NS_{T_{i}}$ is determined as Equation (6), where CNP$_\nu$ is a CNP that is a subset of $T_{i}$.
\begin{equation}
 NS_{T_{i}} = \sum\nolimits_{\text{CNP}_\nu \subseteq T_i} (\text{freq}(\text{CNP}_\nu) \times \text{len}(\text{CNP}_\nu)^2)
\end{equation}
\item Anomaly score calculation. \\
Similarly, an anomaly score $AS_{T_{i}}$ is computed for each test instance if a CAP is its subset, which is determined by Equation (7) and CAP$_\omega$ is a CAP that is a subset of $T_{i}$.
\begin{equation}
 AS_{T_{i}} = \sum\nolimits_{\text{CAP}_\omega \subseteq T_i} (\text{freq}(\text{CAP}_\omega) \times \text{len}(\text{CAP}_\omega)^2)
\end{equation}
\item Classification regulations. \\
The classification of an instance into normal or anomalous categories is determined based on three regulations:
\begin{enumerate}[\text{Regulation} 1:]
    \item If the anomaly-score is greater than or equal to the normal-score, classify the instance as anomalous; otherwise, it is normal;
    \item If both anomaly-score and normal-score are zero, classify the instance as anomalous;
    \item If the normal-score is less than $\text{NS}_{\text{ave}} - r \times \text{NS}_{\text{std}}$, classify the instance as anomalous, where $r$ is a user-defined parameter, $\text{NS}_{\text{ave}}$ is the average of normal scores, and $\text{NS}_{\text{std}}$ is the standard deviation of normal scores. Regulation 3 tackles the challenge of detecting unknown new attacks during the testing phase, especially when training data lack features of these attacks. Its principal strategy is to identify potential new attacks that, despite having an anomaly-score of zero, exhibit a normal-score considerably lower than the average. It flags instances as potential new attacks when their normal-score falls below a user-defined threshold, set at a specific number of standard deviations from the mean. This user-determined threshold, based on empirical rules, enhances the detection of novel attack types not encountered in the training stage.
\end{enumerate}
\end{enumerate}
Here we provide a calculation example. Suppose we have four instances for a task of test: $T_{\text{1}} = (a, b, c, d, e, f)$, $T_{\text{2}} = (b, c, d, f)$, $T_{\text{3}} = (a, b, d, g, f)$, and $T_{\text{4}} = (s, t, u, \nu)$. After executing step 1 and step 2, we obtain $NS_{T_{\text{1}}} = 13$ (3×2 from $(a, b, e)$, 4×1 from $(a, b, d, e)$, and 3×1 from $(a, b, f)$), $AS_{T_{\text{1}}} = 3$ (3×1 from $(b, c, d)$), $NS_{T_{\text{2}}} = 0$, $AS_{T_{\text{2}}} = 3$, $NS_{T_{\text{3}}} = 3$, $AS_{T_{\text{3}}} = 3$, $NS_{T_{\text{4}}} = 0$, and $AS_{T_{\text{4}}} = 0$. Step 3 produces the result of classification as follow. $T_{\text{1}}$ is normal by Regulation 1. $T_{\text{2}}$ is anomaly by Regulation 1. $T_{\text{3}}$ is anomaly because of the following reason. Initially, by Regulation 1, $T_{\text{3}}$ is normal. But, by Regulation 3, given $r$ = 0.1, $NS_{T_{\text{3}}} < \text{NS}_{\text{ave}} - r \times \text{NS}_{\text{std}}$ as well as $3 < 4 - (0.1 \times 6.164)$ , and hence $T_{\text{3}}$ is anomaly eventually. $T_{\text{4}}$ is anomaly by Regulation 2.
\begin{algorithm}
    \caption{Score by patterns} 
    \begin{algorithmic}[1]
        \Statex \textbf{Input:} Instances of test T, distinguishable patterns and score
        \Statex \textbf{Output:} Normal score (NS), Abnormal score (AS)
        \For {$\text{each instance} \; T_i \in T$}
            \State $NS_{T_i} \leftarrow 0,\, AS_{T_i} \leftarrow 0$
            \For {$\text{each pattern}\; \nu \in \text{CNP}$} 
                \If {$\nu \subseteq T_i$}
                    \State $NS_{T_i} \leftarrow NS_{T_i} + \text{freq}(\text{CNP}_{\nu})\times \text{len}(\text{CNP}_{\nu})^2$
                \EndIf
            \EndFor
            \For {$\text{each pattern}\; \omega \in \text{CAP}$}
                \If {$\omega \subseteq T_i$}
                    \State $AS_{T_i} \leftarrow AS_{T_i} + \text{freq}(\text{CAP}_{\omega})\times \text{len}(\text{CAP}_{\omega})^2$
                \EndIf
            \EndFor
        \EndFor
        \State $\text{NS}_\text{{ave}} \leftarrow \textsc{Ave}(\text{NS})$ 
        \State $\text{NS}_\text{{std}} \leftarrow \textsc{Std}(\text{NS})$ 
        \For {$\text{each instance} \; T_i \in T$}
            \If {$AS_{T_i} \geq NS_{T_i}$}
                \State Classify $T_i$ as "Anomaly"
            \ElsIf {$NS_{T_i} < \text{NS}_\text{{ave}} - r \times \text{NS}_\text{{std}}$}
                \State Classify $T_i$ as "Anomaly"
            \Else
                \State Classify $T_i$ as "Normal"
            \EndIf
        \EndFor
    \end{algorithmic} 
\end{algorithm}

\subsection{Information Forensics}
IG possesses the ability of knowledge discovery which helps cybersecurity experts in comprehending the underlying causes of attacks. The study \cite{linkov2020cybertrust} has explored what attack is associated with a single feature, however we have proven the single feature exists on normal instances as well. IG distinguish coherent patterns through set theory. Usually both normal and attack patterns exist simultaneously. Set theory proves that CAP are mutually exclusive from normal traffic patterns, which eliminates a contradiction between anomalous and normal patterns. Without uncertainty, IG maps coherent patterns onto comprehensive intrusion paths. Further, different intrusion paths share commonality, which is helpful in understanding essentials of attacks. Specifically, the commonality provides not only frequency of the attacks but also reflection of cybersecurity vulnerability. In the experiment, we show the most frequent attack intrusion paths at incremental ratios of training to test.
\section{Experimentation}
\subsection{Datasets Description}
We conducted experiments with three well-known public datasets NSL-KDD \cite{tavallaee2009detailed}, UNSW-NB15 \cite{moustafa2015unsw}, and UKM-IDS20 \cite{al2021adaptive} as detailed in Table \ref{tab:table2}. The NSL-KDD dataset, a refined version of the original KDD99 dataset, was developed by Tavallaee et al. to address certain limitations of KDD99. This refinement involved creating various subsets, namely KDDTrain+, KDDTest+, KDDTrain\_20Percent, and KDDTest-21. NSL-KDD is distinguished by the removal of redundant records from KDD99 and the introduction of a metric called \#successfulPrediction, derived from the results of 21 different experiments. Consequently, 125,973 records were selected for KDDTrain+ and 22,544 for KDDTest+, based on the success rate indicated by the successful prediction metric. The KDDTrain\_20Percent subset comprises the initial 20\% of the KDDTrain+ dataset. KDDTest-21, another refined subset from KDDTest+, was created by excluding records that were consistently and correctly classified in all 21 experiments. NSL-KDD categorizes the anomalous records into four primary attack types: Denial of Service (DoS), Probe, Remote to Local (R2L), and User to Root (U2R). Regarding features, the NSL-KDD dataset builds upon the original 42 features of KDD99 by adding an additional column, which records the frequency of correct classifications in the 21 experiments, thereby providing an additional dimension for analysis.

\begin{table}
 \caption{Complete Dataset Counts of Normal and Attack Instances for NSL-KDD, UNSW-NB15, UKM-IDS20}
  \centering
  \begin{tabular}{lrrr}
    \toprule
    Dataset     & Normal     & Anomaly     & Total\\
    \midrule
    NSLKDD -KDDTrain+ & 67,343  & 58,630 & 125,973 \\
    NSLKDD -KDDTrain\_20Percent & 13,449  & 11,743 & 25,192 \\
    NSLKDD -KDDTest+ & 9,711  & 12,833 & 22,544 \\
    NSLKDD -KDDTest-21 & 2,152  & 9,698 & 11,850 \\
    \hline
    UNSW-NB15 -Training & 56,000  & 119,341 & 175,341 \\
    UNSW-NB15 -Test set & 37,000  & 45,332 & 82,332 \\
    \hline
    UKM-IDS20 -Training & 7,140  & 3,168 & 10,308 \\
    UKM-IDS20 -Test set & 1,769  & 810 & 2,579 \\
    \bottomrule
  \end{tabular}
  \label{tab:table2}
\end{table}
The UNSW-NB15 dataset \cite{tavallaee2009detailed} consists of four CSV files, which together contain 2,540,044 records. This dataset is characterized by its diversity, comprising 49 unique features and covering nine types of attacks: Worms, Shellcode, Reconnaissance, Generic, Fuzzers, Exploits, DoS, Backdoor, and Analysis. These features are classified into various categories, including basic, content, temporal, and flow features, in addition to several derived features. Each record in the dataset is annotated with labels for both attack type and binary classification. Due to the considerable size of the UNSW-NB15 dataset, it has been divided into separate subsets for training and testing purposes.

The UKM-IDS20 dataset, developed by Al-Daweri et al. \cite{al2021adaptive}, was created by collecting original network traffic data from the National University of Malaysia (UKM). This effort was undertaken to simulate attack activities in a realistic environment, with the data collection spanning over a period of two weeks. The dataset includes 48 distinct features, encompassing an attack type label, a binary classification label, basic features, connection flag features, connection count features, size-based features, and time-based features. This sentence was already clear and concise, accurately describing the types of attacks covered by the dataset. Compared to benchmark datasets, UKM-IDS20 provides valuable insights into contemporary network traffic and modern attack methods.
\subsection{Data Preprocessing}
In this research, limitations due to large-scale data necessitated extracting only portions of the datasets for our experiments. To ensure reproducibility, instances were selected sequentially instead of randomly, as detailed in Table \ref{tab:table3}. In NSL-KDD, the first 10,000 instances from the Train+ subset and the first 5,000 instances from the Test+ subset were utilized. In UNSW-NB15, a total of 14,714 instances were extracted from its four CSV files. These anomalous instances were evenly distributed across nine attack categories, with each category contributing 500 instances. Where an attack category had fewer than 500 instances, all available instances were included. In UKM-IDS20, the entire dataset from both training and testing subsets was used. Table \ref{tab:table4} presents the proportion of each attack type within these datasets.

\begin{table}
 \caption{Experimental Subset Counts of Normal and Attack Instances for NSL-KDD, UNSW-NB15, UKM-IDS20}
  \centering
  \begin{tabular}{rrrr}
    \toprule
    Dataset     & Normal     & Anomaly     & Total\\
    \midrule
    NSL-KDD & 7,439  & 7,561 & 15,000 \\
    UNSW-NB15 & 10,000  & 4,174 & 14,174 \\
    UKM-IDS20 & 8,909  & 3,978 & 12,887 \\
    \bottomrule
  \end{tabular}
  \label{tab:table3}
\end{table}

\begin{table}
 \caption{The proportion of attacks in each dataset}
  \centering
  \begin{tabular}{p{2.5cm}p{9.5cm}}
    \toprule
    Dataset     & Attacks: (Type, Amounts)\\
    \midrule
    NSL-KDD & (DoS, 5388), (Probe, 1452), (R2L, 669), (U2R, 52) \\
    UNSW-NB15 & (Analysis, 500), (Backdoors, 500), (DoS, 500),
    (Exploits, 500), (Fuzzers, 500), (Generic, 500),
    (Reconnaissance, 500), (Shellcode, 500), (Worms, 174) \\
    UKM-IDS20 & (ARP poisoning, 592), (DoS, 1742)
    (Scans, 597), (Exploits, 1047) \\
    \bottomrule
  \end{tabular}
  \label{tab:table4}
\end{table}

In the data preprocessing phase, instances with identical features but differing class labels were excluded to avoid contradictions. For the UNSW-NB15 and UKM-IDS20 datasets, no contradictory instances were identified, whereas in the NSL-KDD dataset, 133 instances required exclusion. The adjusted total number of instances is presented in Table \ref{tab:table5}, which depicts the '1|9' training-to-testing ratio as 10\%-to-90\%. Moreover, Table \ref{tab:table5} details the distribution of normal and anomalous instances within the training and testing sets, emphasizing the variations in these distributions at various training-to-testing ratios.
\subsection{Result}
Evaluating the performance of predictive models is essential, especially in the context of IDS, where accurately distinguishing between normal and anomalous activities is critical. The Confusion Matrix, a fundamental tool in classification problems, plays a vital role in our experiments. In the proposed IG, the "anomaly" class is labeled as positive, while the "normal" class is negative. Four aspects of this matrix were analyzed such as True Positive (TP), False Positive (FP), True Negative (TN), and False Negative (FN) rates. TP measures how well the system identifies actual anomalies, FP indicates normal instances incorrectly labeled as anomalies, TN represents correct predictions of normal instances, and FN accounts for anomalies incorrectly labeled as normal.

To thoroughly assess the capability of IG, we use the key metrics: accuracy, precision, recall, and the Area Under the Receiver Operating Characteristic Curve (AUC). Accuracy is the proportion of all predictions (both anomalies and normal) that a method correctly identifies, reflecting overall effectiveness. However, due to the imbalanced distribution between normal and anomaly classes in cybersecurity data, precision and recall are critical for a more detailed evaluation. Precision is the ratio of correctly predicted anomaly instances to all instances predicted as anomalies, essentially measuring the exactness of a method in identifying true threats. In contrast, recall, also known as the sensitivity, assesses the capability of a method to detect actual anomalies from all true anomaly instances.

The AUC provides a comprehensive measure of a method's capability to distinguish between actual attacks and normal activities across various decision-making points. Unlike precision and recall, which assess a method's performance at a single, specific point (like a fixed level of alertness), AUC evaluates a method's effectiveness across a spectrum of conditions, offering a broader view of its overall performance. This range of conditions could represent different levels of strictness or sensitivity in identifying threats, making AUC a crucial metric in assessing IDS effectiveness, especially given the diverse nature of cyber threats. Table 6 presents the performance of IG as:

\begin{table}
\tiny
 \caption{Quantitative Ratios of Normal and Anomalous Instances (Excluded Contradictory Instances)}
  \centering
  \begin{tabular}{c|rrrr|rrrr|rrrr}
    \toprule
    \multirow{3}{*}{Ratios}     & \multicolumn{4}{c|}{NSL-KDD}     & \multicolumn{4}{c|}{UNSW-NB15}     & \multicolumn{4}{c}{UKM-IDS20}\\
    \cline{2-13}    
    & \multicolumn{2}{c}{Training}    & \multicolumn{2}{c|}{Test}         & \multicolumn{2}{c}{Training}    & \multicolumn{2}{c|}{Test}         & \multicolumn{2}{c}{Training}    & \multicolumn{2}{c}{Test} \\
    \cline{2-13}
    & \multicolumn{1}{r}{Anomaly} & Normal                  & Anomaly & Normal                  & Anomaly & Normal                  & Anomaly & Normal                  & Anomaly & Normal                  & Anomaly & Normal \\
    \hline
    1 | 9 & 743 & 744 & 6726 & 6654 & 417 & 1000 & 3757 & 9000 & 409 & 880 & 3569 & 8029 \\
    2 | 8 & 1418 & 1555 & 6051 & 5843 & 835 & 2000 & 3339 & 8000 & 848 & 17 29 & 3130 & 7180 \\
    3 | 7 & 2127 & 2333 & 5342 & 5065 & 1252 & 3000 & 2922 & 7000 & 1237 & 2629 & 2741 & 6280 \\
    4 | 6 & 2784 & 3163 & 4685 & 4235 & 1670 & 4000 & 2504 & 6000 & 1628 & 3527 & 2350 & 5382 \\
    5 | 5 & 3482 & 3952 & 3987 & 3446 & 2087 & 5000 & 2087 & 5000 & 2006 & 4438 & 1972 & 4471 \\
    6 | 4 & 4174 & 4746 & 3295 & 2652 & 2504 & 6000 & 1670 & 4000 & 2398 & 5334 & 1580 & 3575 \\
    7 | 3 & 4932 & 5475 & 2537 & 1923 & 2922 & 7000 & 1252 & 3000 & 2787 & 6234 & 1191 & 2675 \\
    8 | 2 & 5757 & 6137 & 1712 & 1261 & 3339 & 8000 & 835 & 2000 & 3169 & 7141 & 809 & 1768 \\
    9 | 1 & 6611 & 6769 & 858 & 629 & 3757 & 9000 & 417 & 1000 & 3585 & 8013 & 393 & 896 \\ 
    \bottomrule
  \end{tabular}
  \label{tab:table5}
\end{table}

\begin{table}
\tiny
 \caption{Performance Metrics Across Quantitative Ratios for NSL-KDD, UNSW-NB15, and UKM-IDS20}
  \centering
  \begin{tabular}{c|rrrr|rrrr|rrrr}
    \toprule
    \multirow{2}{*}{Ratios}     & \multicolumn{4}{c|}{NSL-KDD (std=0.55)}     & \multicolumn{4}{c|}{UNSW-NB15 (std=0.86)}     & \multicolumn{4}{c}{UKM-IDS20 (std=0.82)}\\
    \cline{2-13}    
    & Accuracy & Recall                  & Precision & AUC                  & Accuracy & Recall                  & Precision & AUC                  & Accuracy & Recall                  & Precision & AUC                   \\
    \hline
    1 | 9 & 0.9389 & 0.9353 & 0.9426 & 0.9402 & 0.9894 & 0.9997 & 0.9656 & 0.9988 & 0.9810 & 0.9980 & 0.9436 & 0.9982 \\
    2 | 8 & 0.9351 & 0.9341 & 0.9381 & 0.9276 & 0.9923 & 0.9994 & 0.9752 & 0.9991 & 0.9709 & 1.0000 & 0.9125 & 0.9995 \\
    3 | 7 & 0.9108 & 0.8437 & 0.9798 & 0.9291 & 0.9941 & 0.9997 & 0.9805 & 0.9995 & 0.9605 & 1.0000 & 0.8851 & 0.9996 \\
    4 | 6 & 0.9325 & 0.9276 & 0.9429 & 0.9111 & 0.9953 & 1.0000 & 0.9843 & 0.9996 & 0.9631 & 1.0000 & 0.8918 & 0.9999 \\
    5 | 5 & 0.9296 & 0.9175 & 0.9496 & 0.8952 & 0.9949 & 1.0000 & 0.9830 & 0.9996 & 0.9601 & 1.0000 & 0.8847 & 0.9999 \\
    6 | 4 & 0.9171 & 0.9062 & 0.9420 & 0.8709 & 0.9954 & 1.0000 & 0.9847 & 0.9997 & 0.9595 & 1.0000 & 0.8832 & 0.9999 \\
    7 | 3 & 0.9435 & 0.9752 & 0.9290 & 0.9602 & 0.9951 & 1.0000 & 0.9835 & 0.9998 & 0.9571 & 1.0000 & 0.8777 & 0.9998 \\
    8 | 2 & 0.9381 & 0.9889 & 0.9112 & 0.9871 & 0.9954 & 1.0000 & 0.9847 & 0.9998 & 0.9631 & 1.0000 & 0.8949 & 0.9998 \\
    9 | 1 & 0.9671 & 0.9650 & 0.9776 & 0.9931 & 0.9944 & 1.0000 & 0.9812 & 0.9998 & 0.9620 & 1.0000 & 0.8891 & 0.9999 \\ 
    \bottomrule
  \end{tabular}
  \label{tab:table6}
\end{table}

\begin{table}
\tiny
 \caption{Performance comparison of Methods in NSL-KDD, UNSW-NB15, and UKM-IDS20}
  \centering
  \begin{tabular}{c|rrrr|rrrr|rrrr}
    \toprule
    \multirow{2}{*}{Methods}     & \multicolumn{4}{c|}{NSL-KDD}     & \multicolumn{4}{c|}{UNSW-NB15}     & \multicolumn{4}{c}{UKM-IDS20}\\
    \cline{2-13}    
    & Accuracy & Recall                  & Precision & AUC                  & Accuracy & Recall                  & Precision & AUC                  & Accuracy & Recall                  & Precision & AUC                   \\
    \hline
    \cite{gu2021effective} & 99.35 & N/A & N/A & N/A & 93.75 & N/A & N/A & N/A & \multicolumn{4}{c}{N/A}\\
    \cite{chiba2019intelligent} & 99.86 & 99.93 & 99.83 & 99.87 & \multicolumn{4}{c|}{N/A} & \multicolumn{4}{c}{N/A}\\
    \cite{gottwalt2019corrcorr} & \multicolumn{4}{c|}{N/A} & 98.65 & N/A & 99.74 & 97 & \multicolumn{4}{c}{N/A}\\
    \cite{patil2019designing} & \multicolumn{4}{c|}{N/A} & 97.09 & N/A & N/A & N/A & \multicolumn{4}{c}{N/A}\\
    \cite{zhang2020model} & \multicolumn{4}{c|}{N/A} & 95.6 & N/A & N/A & N/A & \multicolumn{4}{c}{N/A}\\
    \cite{mohammad2020improved} & \multicolumn{4}{c|}{N/A} & 93.9 & 94.5 & 93.3 & N/A & \multicolumn{4}{c}{N/A}\\
    \cite{aksu2022mga} & \multicolumn{4}{c|}{N/A} & 1 & 1 & 1 & N/A & \multicolumn{4}{c}{N/A}\\
    \cite{wang2021intrusion} & 76.64 & 95.98 & 61.55 & N/A & 72.38 & 69.94 & 87.42 & N/A & \multicolumn{4}{c}{N/A}\\
    \cite{al2021adaptive} & \multicolumn{4}{c|}{N/A} & \multicolumn{4}{c|}{N/A} & 94.66 & 96.91 & 92.43 & N/A  \\
    \hline
    IG (1|9) & 93.89 & 94.26 & 93.53 & 94.02 & 98.94 & 96.56 & 99.97 & 99.88 & 98.10 & 94.36 & 99.80 & 99.82 \\ 
    \bottomrule
  \end{tabular}
  \label{tab:table7}
\end{table}

\begin{enumerate}
    \item \textit{Generalization Across Diverse Scenarios:} IG’s generalization capabilities are exemplified by its consistent performance across datasets: achieving accuracies of 0.9389 to 0.9476 in NSL-KDD, 0.9690 to 0.9782 in UNSW-NB15, and 0.9554 to 0.9611 in UKM-IDS20, across varying training-to-test ratios. These results highlight its robust adaptability to different network environments.
    \item \textit{Precision in Anomaly Detection:} Precision, a critical metric for reducing false alarms in intrusion detection, is a standout feature of IG. In the UKM-IDS20 dataset, even with a training ratio as low as 20\%, IG achieves a precision rate of at least 0.88. This high precision underlines IG’s effectiveness in accurately detecting anomalies, thereby minimizing the occurrence of false positives.
    \item \textit{Efficiency in Identifying True Threats:} IG consistently achieves high recall rates across datasets, exemplified by rates of 1.0 in UNSW-NB15 and UKM-IDS20, and ranging from 0.9353 to 0.9476 in NSL-KDD. These results highlight its capability to accurately identify genuine threats, ensuring minimal misses in detecting actual attacks.
    \item \textit{Comprehensive Performance Evaluation with AUC:} The AUC scores of IG across different datasets underscore its comprehensive effectiveness in distinguishing between normal activities and actual threats. In NSL-KDD, IG achieves an AUC range of 0.9402 to 0.9504, with its peak at 0.9504, demonstrating robust discrimination capability. In UNSW-NB15, the AUC scores hover around a high of 0.99, reflecting exceptional accuracy in threat detection. Similarly, in UKM-IDS20, IG consistently shows AUC scores at 0.99, indicating its sustained effectiveness across various scenarios.
\end{enumerate}

Table \ref{tab:table7}  offers a rigorous comparison of various methods in NSL-KDD, UNSW-NB15, and UKM-IDS20. The performance metrics used for comparison are Accuracy, Recall, Precision, and AUC. 

\begin{figure}
  \centering
  \includegraphics[width=16.5cm]{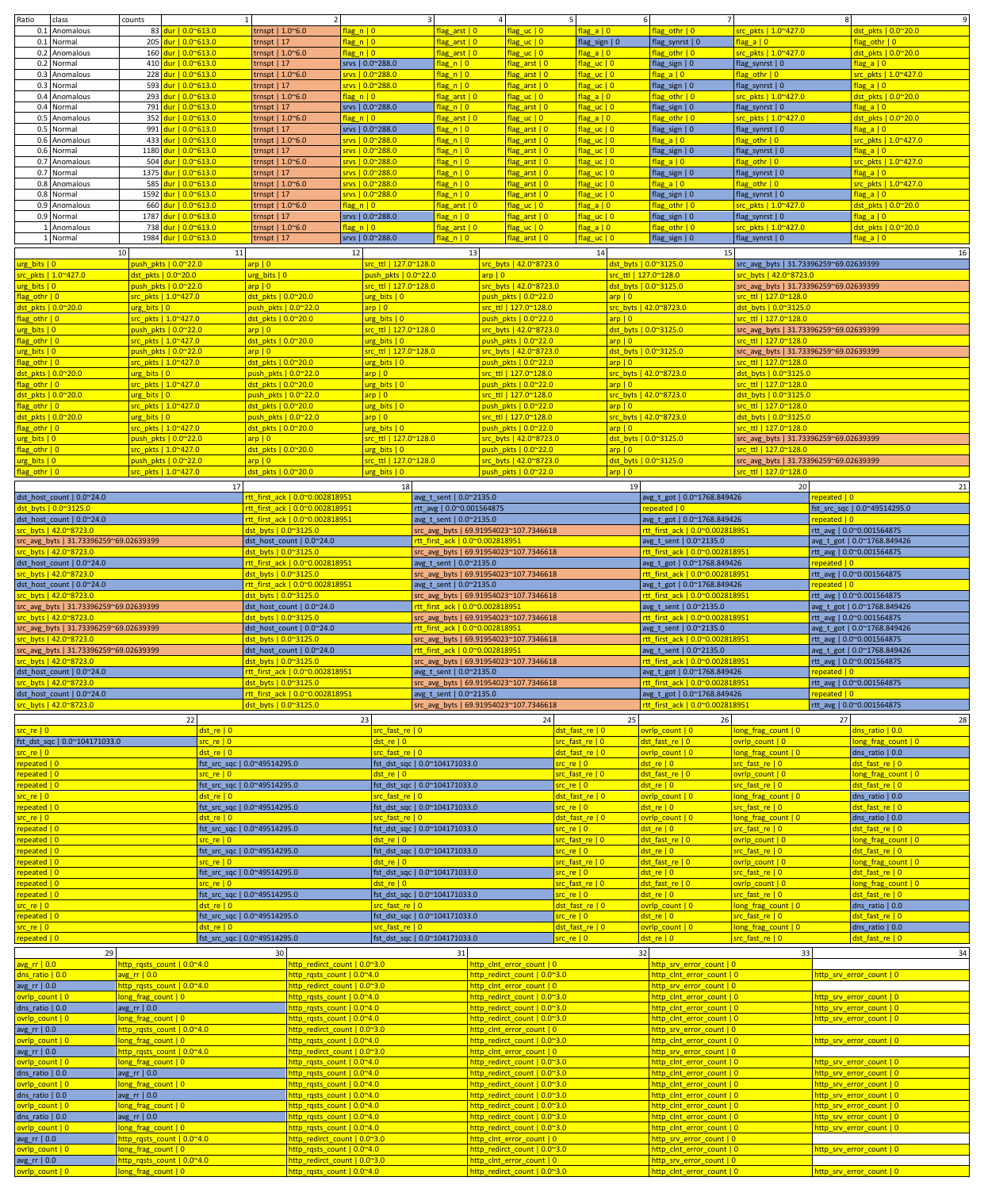}
  \caption{The Most Frequent Intrusion Techniques in UKM-IDS20.}
  \label{fig:fig2}
\end{figure}

\begin{figure}
  \centering
  \includegraphics[width=16.5cm]{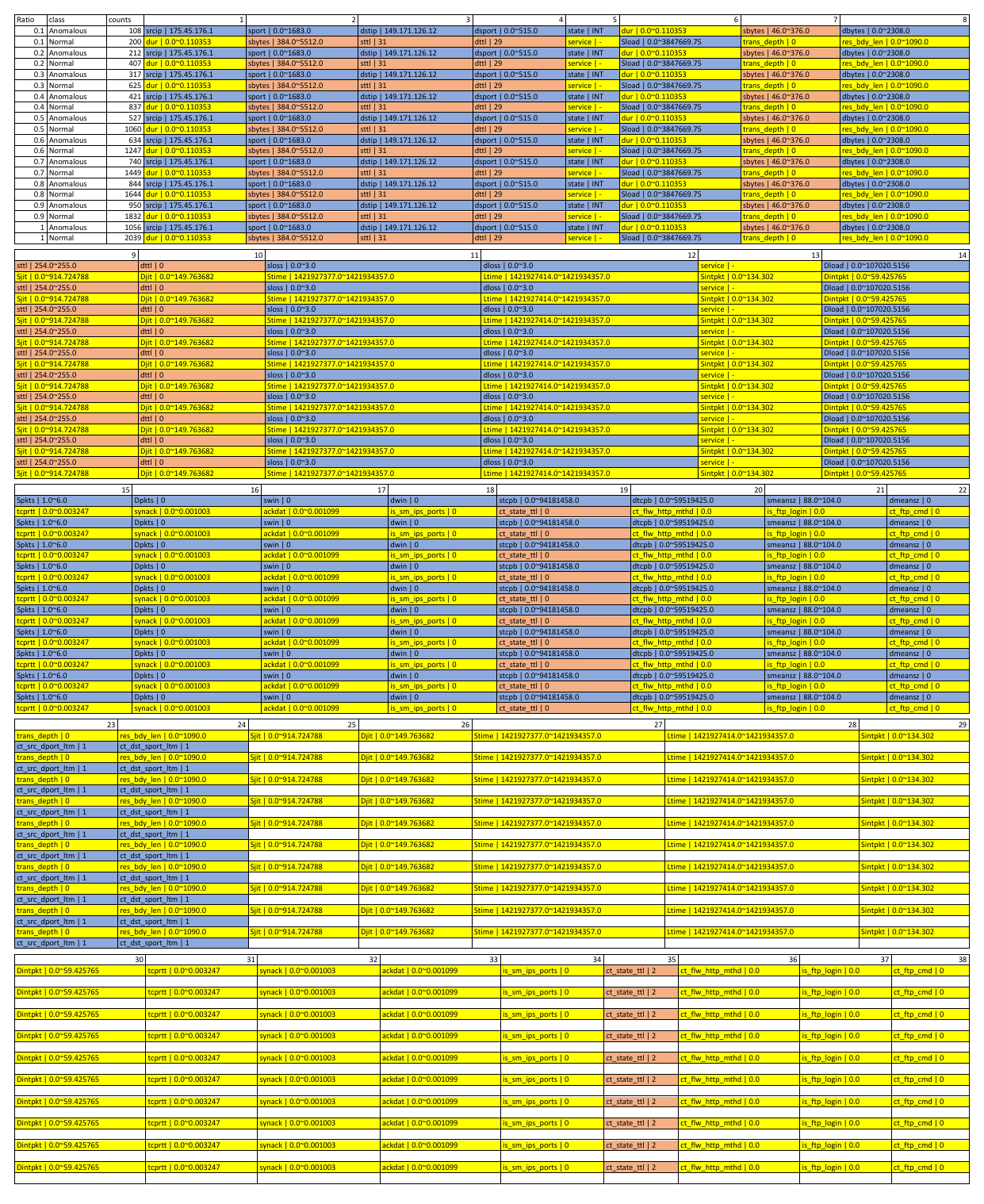}
  \caption{The Most Frequent Intrusion Techniques in UNSW-NB15.}
  \label{fig:fig3}
\end{figure}

\begin{figure}
  \centering
  \includegraphics[width=16.5cm]{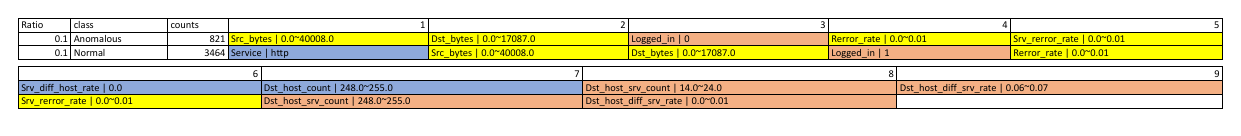}
  \caption{The Most Frequent Intrusion Techniques in NSL-KDD.}
  \label{fig:fig4}
\end{figure}

\begin{enumerate}
    \item \textit{In NSL-KDD Dataset:} Method \cite{chiba2019intelligent} exhibited outstanding performance, achieving the highest metrics with 99.86\% accuracy, 99.93\% recall, 99.83\% precision, and 99.87\% AUC. It's noteworthy that in NSL-KDD, \cite{chiba2019intelligent} achieved the best results but utilized only 12 features out of the original 41, and its training-to-testing ratio was 52\%-48\%. This distinction in feature selection and data split ratio poses considerations for general applicability and robustness against sophisticated attack scenarios. IG, under a 10\%-90\% training-testing ratio, achieved a competitive performance with 93.89\% accuracy, 94.26\% recall, 93.53\% precision, and 94.02\% AUC, highlighting its robustness even with limited training data.
    \item \textit{In UNSW-NB15 Dataset:} Method \cite{aksu2022mga} reached a perfect score across all metrics, setting a high benchmark. However, it’s noteworthy that the top performance of the method \cite{aksu2022mga} used just 7 of 47 original features at a 50\%-50\% training-testing split. This reduction in feature space, while effective, may limit the model's ability to recognize diverse attack vectors. IG, under a 10\%-90\% training-testing ratio, showcased exemplary performance with 98.94\% accuracy, 96.56\% recall, 99.97\% precision, and 99.88\% AUC. These results are particularly significant, reflecting IG’s superior capability in accurately detecting anomalies with minimal false positives.
    \item \textit{In UKM-IDS20 Dataset:} Method \cite{moustafa2015unsw} displayed a strong performance with 94.66\% accuracy, 96.91\% recall, and 92.43\% precision. IG stood out with its high scores of 98.10\% accuracy, 94.36\% recall, 99.80\% precision, and 99.82\% AUC, underscoring its effectiveness in diverse operational conditions.
\end{enumerate}
As illustrated in Fig. \ref{fig:fig2}, Fig. \ref{fig:fig3}, and Fig. \ref{fig:fig4}, IG methodically discovers coherent patterns, delineating them as digital footprints that sharply distinguish between anomalous and normal network behaviors. These patterns, represented in distinct colors, serve as pivotal evidence in our analysis: blue for features unique to either anomalous or normal behaviors, green for identical features, and red for commands shared between both but with differing parameters. This nuanced identification underscores IG’s comprehensiveness to threat detection, leveraging the full feature set and offering a significant advancement over models restricted by a limited feature scope. Not only does this method demonstrate IG’s robust capability in detecting a wide spectrum of cyber threats, but it also equips cybersecurity professionals with a precise forensic toolkit. Such detailed pattern recognition, devoid of ambiguity, transforms IG into an indispensable asset for both intrusion detection and the meticulous scrutiny integral to digital forensics, highlighting its adaptability and precision in a real-world security landscape.

\section{Conclusion}
In this study, we proposed an interpretable generalization mechanism (IG) for intrusion detection systems (IDS), aiming to not only enhance the performance of intrusion detection by effectively detecting anomalies but also contribute to provide comprehensive intrusion paths and interpretability for the decision-making process. The coherent patterns provide interpretability in differentiating between normal and anomalous traffic, which enables experts to proactively prevent attacks rather than passively detecting them.

IG has primarily five advantages. First, \textit{Interpretability}, IG relies on coherent patterns to differentiate between normal and anomalous traffic, shedding light on the reasoning behind its decisions. Second, \textit{Forensics}, combinations of coherent patterns provide deep insights to grasp comprehensive intrusion paths, facilitating development of cybersecurity. Previously, intrusion paths with partial patterns were indicated as certain type of attacks; however, we discovered the paths also frequently appear in normal traffic so that vast amounts of false alarms arise. In IG, coherent patterns prove comprehensive intrusion paths belong to anomalous traffic only. Third, \textit{Reproducibility}, the entire process of IG is reproducible, which includes data preparation, training, test, evaluation, and inference. IG yields reliable results in both experimental settings and inference scenarios. Fourth, \textit{Effectiveness}, IG is accurate in identifying both normal and anomalous traffic in the real-world public datasets. In Particular, IG achieves high value of AUC even when the proportion of training to test is low, such as AUC=0.94 and Training/Test=10\%/90\% in NSL-KDD, AUC=0.99 and Training/Test=10\%/90\% in UNSW-NB15, AUC=0.99 and Training/Test=10\%/90\% in UKM-IDS20. Fifth, \textit{Generalization}, IG is qualified because it possesses the three capabilities: high PRE, REC, and AUC across various datasets and ratios; identification of anomalous instances without training inclusion; and reproducibility of results. In sum, IG paves the way for more advanced research and development in the realm of explainable and reliable AI-driven security solutions.

%Bibliography
\bibliographystyle{unsrt}  
\bibliography{references}

\end{document}